\documentclass[prd,twocolumn,amsmath,amssymb,floatfix,superscriptaddress]{revtex4-1}

\usepackage{graphicx}
\usepackage{amssymb}
\usepackage{amsmath}
\usepackage{hyperref}

\usepackage{color}

\DeclareMathAlphabet\mathbfcal{OMS}{cmsy}{b}{n}
\definecolor{darkgreen}{cmyk}{0.85,0.2,1.00,0.2} 
\definecolor{purple}{cmyk}{0.5,1.0,0,0}

\def\barray{\begin{array}} 
\def\earray{\end{array}}
\def\be{\begin{equation}}
\def\ee{\end{equation}}
\def\ben{\begin{equation} \nonumber}
\def\een{\end{equation}}
\def\ban{\begin{eqnarray*}}
\def\ean{\end{eqnarray*}}
\def\ba{\begin{eqnarray}}
\def\ea{\end{eqnarray}}

\def\({\left(}
\def\){\right)}

\newcommand{\tr}[1]{[#1]}

\newcommand{\ul}[3]{#1^{#2}_{\hphantom{#2}#3}}
\newcommand{\lu}[3]{#1_{#2}^{\hphantom{#2}#3}}
\newcommand{\fid}{\Sigma}
\newcommand{\bgamma}{\boldsymbol{\gamma}}
\newcommand{\bfid}{\boldsymbol{\Sigma}}
\newcommand{\bg}{{\bf g}}
\newcommand{\bmin}{\boldsymbol{\eta}}   
\newcommand{\blambda}{\boldsymbol{\Lambda}}   
\newcommand{\bvier}{{\bf e}}
\newcommand{\bfvier}{{\bf L}}

\newcommand{\stucky}{St\"{u}ckel\-berg}
\newcommand{\af}{a}
\newcommand{\afd}{\dot a}

\begin{document}

\title{Self-accelerating Massive Gravity:\\ How Zweibeins Walk through Determinant Singularities}
\author{Pierre Gratia}
\email{pgratia@uchicago.edu}
\affiliation{Department of Physics, University of Chicago, Chicago, Illinois 60637, U.S.A}
\author{Wayne Hu}
\email{whu@background.uchicago.edu}
\author{Mark Wyman}
\email{markwy@oddjob.uchicago.edu}
\affiliation{Kavli Institute for Cosmological Physics, Department of Astronomy \& Astrophysics,  Enrico Fermi Institute, University of Chicago, Chicago, Illinois 60637, U.S.A}
\begin{abstract}
The theory of massive gravity possesses ambiguities when the
spacetime metric evolves far from the non-dynamical fiducial metric used to define it.   We explicitly
construct a spherically symmetric example case where the metric evolves to a coordinate-independent determinant singularity which does not
exist in the initial conditions.   Both the metric and the vierbein formulation of the
theory are ill-defined at this point.  In unitary gauge, the chart of the spacetime ends at this point
and does not cover the full spacetime whereas the spherically symmetric vierbeins, or zweibeins, of the fiducial metric become non-invertible
and do not describe a valid metric.   
Nonetheless it is possible to continuously join a zweibein solution on the other side of the singularity which picks one of the degenerate solutions of the metric square root.   This continuous solution is not the choice conventionally made in the previous literature.  We also show that the \stucky\ equations of motion on the self-accelerating branch prevent solutions from evolving to  a more pathological situation in which the spacetime vierbeins lack a crucial symmetry with the fiducial vierbeins and real square roots fail to exist.
\end{abstract}

\maketitle
\section{Introduction}

Massive gravity is a theory with two metrics.   At the linearized level, the Fierz-Pauli theory represents five polarizations of a massive graviton propagating on a flat fiducial background
\cite{Fierz:1939ix}.  
The nonlinear generalization of this theory that preserves the absence of the sixth mode, the
Boulware-Deser ghost \cite{Boulware:1972zf}, retains the concept of a second fiducial metric
 \cite{deRham:2010ik,deRham:2010kj,Hassan:2011hr}.  
  In the simplest version which we treat here, only the spacetime
metric is dynamical whereas the second metric is taken to be Minkowski.
When the spacetime metric evolves to a point where it deviates far from the Minkowski 
fiducial metric, massive gravity enters a new regime where potential pathologies or ambiguities can arise.

A fixed fiducial metric breaks diffeomorphism invariance.  In our case the fiducial
metric is only Minkowski for a specific choice of spacetime coordinates called unitary gauge.
Its representation in a general coordinate system is carried by the Jacobian transform,
also known as the \stucky\ fields or the vierbeins of the fiducial metric.
   
Many of the novel properties of massive gravity arise from this relationship between
the fiducial metric and the spacetime metric (e.g.~\cite{Deffayet:2011rh,Volkov:2013roa}).   For example, since the fiducial metric
is static in unitary gauge, it is not generally possible to accommodate 
the usual homogeneous and isotropic expanding spacetime as a solution \cite{D'Amico:2011jj,Gumrukcuoglu:2011ew,Gumrukcuoglu:2011ew}.  This does not
mean that the Friedmann-Robertson-Walker (FRW) metric is incompatible with massive gravity,
but rather that in these isotropic coordinates the fiducial metric is not Minkowski.
Or, equivalently, in the coordinate system where the fiducial metric is Minkowski, the spacetime
metric is not in its standard homogeneous and isotropic form
\cite{Koyama:2011xz,Koyama:2011yg,Nieuwenhuizen:2011sq,Berezhiani:2011mt}. Indeed, in previous work we showed that
for any isotropic matter distribution, massive gravity admits a solution where the mass term acts
as a cosmological constant \cite{Gratia:2012wt} (see also \cite{Kobayashi:2012fz,Volkov:2012cf,Volkov:2012zb}).  

These self-accelerating solutions have been constructed by considering the \stucky\ fields
as dynamical variables and solving the resulting system of equations
\cite{Wyman:2012iw}.   In certain cases,
these fields evolve to a point where the fiducial metric in isotropic coordinates reaches
a determinant singularity which is not present in the initial conditions.  This singularity
is coordinate invariant but manifests itself slightly differently when formulated in unitary coordinates, where the fiducial metric
is Minkowski. When working in that formulation of the theory, the spacetime metric 
itself encounters this singularity and its chart of the spacetime ends at the singular point.
Another unitary chart exists on either side of the singular point, but a single chart cannot cover the
spacetime.
In fact, multiple solutions exist on the other side of the
singularity.   These multiple solutions are related to different possibilities for taking the
square root of the product of the two metrics which leave the metric formulation ill-defined
at these points.
 
 In this paper, we  explore these issues and their possible resolution.   In \S\ref{sec:massive}, we review the construction of massive gravity out of the spacetime and
 fiducial metrics.   We construct an explicit example of a solution that encounters
 a determinant singularity in \S\ref{sec:zerodetall} and discuss the representation of the
 theory in vierbeins or zweibeins in this spherically symmetric case.   In \S\ref{sec:deSitter}
 we examine the opposite case where the curvature
 and coordinate singularities are in the spacetime metric.    We discuss these
 results in \S\ref{sec:discussion}.

\vfill

\section{Massive Gravity}
\label{sec:massive}

The Boulware-Deser ghost free theory of massive gravity is given by the Lagrangian density
 \cite{deRham:2010kj}
\begin{align}
\label{drgt}
\mathcal{L}_G &=\frac{M_{\rm pl}^2}{2}\sqrt{-g}\left[ R-{m^2}\sum_{k=0}^4 \frac{\beta_k}{k!} F_k\left(\sqrt{\bg^{-1} \bfid }\right)\right],
\end{align}
where $M_{\rm pl}=(8\pi G)^{-1}$ is the reduced Planck mass, $R$ is the Ricci scalar, $\bfid$ is the so-called fiducial metric and the $F_k$ terms are
functions of the square root matrix $\bgamma$
\begin{equation}
\ul{\gamma}{\mu}{\alpha} \ul{\gamma}{\alpha}{\nu} = g^{\mu\alpha}\fid_{\alpha\nu}.
\end{equation}
Specifically,
\begin{align}
F_0(\bgamma) & = 1, \nonumber\\
F_1(\bgamma) & = \tr{\bgamma}, \nonumber\\
F_2(\bgamma) & =  \tr{\bgamma}^2 - \tr{\bgamma^2} , \\
F_3(\bgamma) & =\tr{\bgamma}^3 - 3 \tr{\bgamma} \tr{\bgamma^2} + 2 \tr{\bgamma^3} , \nonumber\\
F_4(\bgamma) &= \tr{\bgamma}^4 - 6 \tr{\bgamma}^2 \tr{\bgamma^2} + 3 \tr{\bgamma^2}^2 + 8 \tr{\bgamma} \tr{\bgamma^3}  
- 6 \tr{\bgamma^4} ,
\nonumber
\end{align}
where $[\,]$ denotes the trace of the enclosed matrix.  The parameters of the theory are
$m$, the graviton mass, and $\beta_k$.  Not all of the latter parameters are independent since
\begin{align}
\beta_0 &= -12 (1+ 2\alpha_3+2\alpha_4), \nonumber\\
\beta_1 &= 6(1 + 3 \alpha_3 + 4\alpha_4),\nonumber\\
\beta_2 &= -2(1+ 6 \alpha_3+12\alpha_4 ), \\
\beta_3 &= 6(\alpha_3+ 4\alpha_4), \nonumber\\
\beta_4 &= -24 \alpha_4,\nonumber
\end{align} 
leaving two remaining independent parameters $\{\alpha_3,\alpha_4\}$.

This theory is the nonlinear completion of the Fierz-Pauli action where the spacetime metric
is assumed to be close to the Minkowski metric locally $\fid_{\mu\nu}=\eta_{\mu\nu}$.   Here we wish to examine matter configurations that evolve the spacetime metric
away from the fiducial metric in potentially problematic ways.

The introduction of a Minkowski fiducial metric breaks diffeomorphism invariance.    Working with the theory in this form is equivalent to choosing a preferred set of coordinates, {which is known as}
unitary gauge, to specify the spacetime metric.  We are particularly interested in cases
where this set of coordinates does not cover the whole spacetime.   It is therefore
convenient to restore diffeomorphism invariance with {a set of} \stucky\ fields
and represent the fiducial metric in covariant form
\begin{equation}
\fid_{\mu\nu} = \partial_\mu\phi^a \partial_\nu \phi^b \eta_{a b}.
\label{eqn:stucky}
\end{equation}
The \stucky\ fields $\phi^a$ transform as spacetime scalars.  Throughout, Greek indices denote the spacetime and are lowered and raised with $\bg$ and its inverse; Latin indices likewise by the Minkowski metric $\bmin$. 

Given that if $\tilde x^a = \phi^a$ then $\fid_{\mu\nu}= \eta_{\mu\nu}$, the matrix $\partial_\mu\phi^a$ represents the Jacobian of the coordinate transform between a general set of coordinates
$x^\mu$ and unitary gauge coordinates $\tilde x^\mu$.  When the determinant of this
Jacobian vanishes, this transformation is not invertible nor is the fiducial metric
itself invertible.  
As the notation suggests, the Jacobian matrix $\partial_\mu\phi^a$ is a kind of ``square root" of the fiducial metric, suggesting that it is related to the vierbeins of that metric. We shall use these Jacobian-derived vierbeins to explore the nature of such singular points.

\section{Singular Fiducial Metric}
\label{sec:zerodetall}

We use the explicit solutions developed in Ref.~\cite{Wyman:2012iw} to analyze
cases where the fiducial metric evolves to a determinant singularity that was not
present in the initial conditions.  
In \S\ref{sec:zerodet}, we review the construction of such solutions and their implications
for the unitary gauge chart.   In \S\ref{sec:squareroot}, we relate this point to a degeneracy in 
the possible solutions for the square root that defines the metric formulation of massive gravity.
We discuss the vierbein formulation in \S\ref{sec:vierbeins} and apply it in \S\ref{sec:zweibeins} to match solutions on either side of the determinant singularity.

\subsection{Zero Determinant}
\label{sec:zerodet}

We 
focus on one particular thought example where simple exact solutions exist
\cite{Gumrukcuoglu:2011ew}.   
Here the spacetime metric is an open FRW universe where the solution to the modified 
Einstein equations for massive gravity are identical to Einstein gravity with an effective
cosmological constant for any component of true matter in the background; here and throughout, 
we will use the terminology ``true" matter to denote matter content that exists independently of, as opposed to an effective consequence
of the graviton mass term in Eq.~(\ref{drgt}). 

 {In this example, the singularity
occurs when an initially expanding spacetime ceases to expand and begins to contract
at a finite future time because of the presence of negative stress-energy}.
 To construct
our singularity, we  can either choose
the massive gravity parameters to give a negative effective cosmological constant, e.g.~$\alpha_3=-4\alpha_4$ and $\alpha_4>1/12$, or introduce 
 some more dominant
matter component with negative energy density and $p/\rho<-1/3$, and assume we begin
with initial conditions such that $\dot a>0$ \cite{Wyman:2012iw}.   
In either case the expansion turns to contraction
at some point in time. Although other examples exist where the zero determinant occurs
somewhere in space \cite{Wyman:2012iw},  we construct the example in this way so that no pathologies exist
on some initial time surface in the expansion stage.

Specifically, the open FRW spacetime metric in isotropic coordinates is given by
\begin{equation}
ds^2 = -dt^2 + \left[ \frac{\af(t)}{1+K r^2/4} \right]^2 (dr^2 + r^2 d\Omega^2),
\end{equation}
where the scale factor $\af$ obeys the ordinary Friedmann equation with
spatial curvature $K<0$
\begin{equation}
\left( \frac{\dot \af}{\af} \right)^2 +\frac{K}{\af^2} = \frac{\rho + \rho_{\rm G} }{3M_{\rm pl}^2} .
\end{equation}
Here $\rho$ is the true matter density and
\begin{equation}
\rho_{G} = \frac{1}{2} m^2 M_{\rm pl}^2  P_0(x_0),
\end{equation}
 is the effective
cosmological constant
where 
\begin{equation}
x_0 = \frac{ 1 + 6\alpha_3 + 12\alpha_4 \pm \sqrt{ 1+ 3\alpha_3 + 9\alpha_3^2 - 12 \alpha_4}}{3 (\alpha_3+4\alpha_4)}, \label{gsol}
\end{equation}
and
\begin{align}
P_0(x) &= - 12 - 2 x(x-6) - 12(x-1)(x-2)\alpha_3 
\nonumber\\&\qquad -24(x-1)^2\alpha_4 .
\end{align}
In fact $\rho_{G}$ describes the impact of massive gravity for any isotropic distribution of matter not just that of an open FRW spacetime
\cite{Gratia:2012wt}.  The useful consequence of choosing an open FRW spacetime is
that certain solutions for the fiducial metric are particularly simple.   In terms of the 
\stucky\ fields, \begin{align}
\phi^0 &= f(t,r) ,\nonumber\\
\phi^i &= g(t,r) \frac{x^i}{r} ,
\label{eqn:fg}
\end{align}
one solution is
\begin{align}
f(r,t)& = x_0 \af(t) \sqrt{\frac{1}{-K} + \frac{r^2}{(1+Kr^2/4)^2 }} ,\nonumber\\
g(r,t)& = x_0\af(t) \frac{r}{1+ K r^2/4}  ,
\end{align}
where the fiducial metric is given by
\begin{equation}
g^{\mu\alpha}\fid_{\alpha\nu}=x_0^2 \left(
      \begin{array}{cccc}
        \dfrac{\afd^2}{-K} & 0 & 0 & 0 \\
        0 & 1 & 0 & 0\\
        0 & 0 & 1& 0\\
        0 & 0 & 0 &1
      \end{array} \right).
\end{equation}
From this solution, it is clear that at the moment of turnaround  
\begin{equation}
{\rm det}\left( g^{\mu\alpha}\fid_{\alpha\nu} \right)= \det(\bg^{-1})\det(\bfid) =0, \qquad {\afd=0}.
\end{equation}
Since the spacetime metric has finite determinant, the fiducial metric has a determinant zero singularity at this moment in time everywhere in space.  
Although a coordinate transform can shift the appearance of
the det=0 problem from the fiducial metric to the spacetime metric, since $g^{\mu\alpha}\fid_{\alpha\nu}$ transforms as a tensor with one covariant and one contravariant index its determinant
vanishes in all frames.
Moreover since $\afd>0$ 
on the initial hypersurface, this example shows that singularities
can develop dynamically from initial conditions that lack them.

We can examine the origin of this singularity in terms of unitary gauge coordinates
$\tilde t = \pm f$ and $\tilde r= \pm g$.  In this case, the fiducial metric remains $\eta_{\mu\nu}$ and the singularity is shifted to the spacetime metric.  Here the zero determinant is a consequence of the
vanishing of the Jacobian determinant
\begin{equation}
{\rm det}(\partial_\mu \phi^a)=\dot f g' - \dot g f' = 0,
\end{equation}
and its cause is that unitary time $\tilde t \propto \af(t)$ and so ``stops" at the turnaround.
This ending of unitary coordinates is not necessarily a problem.   For example it is well
known that synchronous coordinates become singular once the initially defined
freely falling trajectories cross.   Even more prosaically, the 2-sphere cannot be covered by a single chart without encountering det=0.   All this means is that one must switch to a new
set of coordinates before these special points.

Analogously we can look
beyond the det=0 point to construct a new unitary gauge.   The difference between this case
and the simple synchronous or 2-sphere examples is that the determinant singularity we are considering
necessarily divides two charts that lack a common range of validity because det=0 is a coordinate invariant.  This lack of mutual validity creates an ambiguity.   There are two choices of how to continue
on the other side of det=0:  
 either the time coordinate continues on 
smoothly as $\tilde t \propto a(t)$ but is double valued in the full spacetime or as
$\tilde t \propto -a(t)$ with two separate charts whose coordinates are not smoothly joined 
at turnaround.  
This choice correspond to  two possible ways to take the square root of
the matrix $g^{\mu\alpha}\fid_{\alpha\nu}$ to find $\ul{\gamma}{\mu}{\nu}$ as we shall
see next.  The consequence of this choice will be considered in \S \ref{sec:squareroot}
and \S \ref{sec:zweibeins}.

\subsection{Square Roots}
\label{sec:squareroot}

In this and other spherically symmetric examples, the task of solving the massive gravity
equation of motion reduces to that of defining the square roots of a 2$\times$2 matrix.   It is therefore
useful to recall the general properties of the square root solutions before examining our
simple thought example.

 The Cayley-Hamilton theorem states that 
a 2$\times$2 matrix ${\bf A}$ satisfies
\begin{equation}
 \tr{{\bf A}} {\bf A} = {\bf A}^2 + (\det{{\bf A}}) \;  {\bf I}_2,
 \label{eqn:CH}
\end{equation}
where ${\bf I}_2$ is the identity matrix.   Since this equation implies
\begin{equation}
{\tr{\bf A}}^2 = \tr{{\bf A}^2} + 2 \det{\bf A},
\label{eqn:traceeqn}
\end{equation}
it also defines ${\bf A}$ in terms of ${\bf A}^2$.
Explicitly, one chooses the sign of the determinant
\begin{equation}
\det{\bf A} = \pm \sqrt{\det{\bf A}^2}
\end{equation}
and then defines 4 possibilities for the trace
\begin{equation}
{\tr{\bf A}} = \pm \sqrt{ \tr{{\bf A}^2} \pm 2 \sqrt{\det{\bf A}^2}}.
\label{eqn:trace}
\end{equation}
For cases where ${\tr{\bf A}}\ne 0$, this leads to 4 solutions for the square root
of ${\bf A}^2$.  When $\det{\bf A}=0$ two pairs of these solutions become degenerate.
When ${\tr{\bf A}}= 0$,
\begin{equation}
 {\bf A}^2 =\mp \sqrt{\det{\bf A}^2} \;  {\bf I}_2,
\end{equation}
and there is a continuum of solutions
\begin{equation}
{\bf A} =
\left( 
\begin{array}{cc}
\alpha & \beta \\
\gamma & -\alpha \\
\end{array} 
\right)
\end{equation}
subject to the constraint
\begin{equation}
\alpha^2 + \beta \gamma = \mp \sqrt{\det{\bf A}^2}.
\end{equation}

Now let us consider the implications for massive gravity.  As a shorthand, we will explicitly only consider the 2$\times$2 time-radius block of matrices since the angular components are trivial.
Choosing initial conditions on some time slice corresponds to choosing one of the 4 
solutions for the square root of $\bg^{-1}\bfid$.    The Friedmann dynamics then evolve the metrics to a point where the determinant vanishes.   Two solutions then become degenerate and we need 
a rule for how to continue the solutions past this point.
  In our thought example, the two
possibilities are
\begin{equation}
\bgamma_{\pm} =x_0
\left( 
\begin{array}{cc}
\pm \dfrac{\afd}{\sqrt{-K}} & 0 \\
0 & 1 \\
\end{array} 
\right)
\label{eqn:gammapm}
\end{equation}
and so coincide at $\afd=0$.

A common convention in the literature \cite{Nieuwenhuizen:2011sq,D'Amico:2011jj,Gumrukcuoglu:2011zh} is to keep det$[\bgamma] >0$ or equivalently to
define positive definite square roots.  This corresponds to taking $\bgamma=\bgamma_+$ for
$\dot a>0$ and $\bgamma = \bgamma_-$ for $\dot a<0$.   In the \stucky\ field equations this is enforced by using a sign parameter
\begin{equation}
\mu = {\rm sgn}(\dot f g' - \dot g f')
\end{equation}
to define the determinant as
\begin{equation}
{\rm det}[\bgamma] \propto \mu (\dot f g' - \dot g f') ,
\label{eqn:mujac}
\end{equation}
which effectively takes the absolute value of the \stucky\ determinant.   
 This has the benefit that given a positive trace and determinant of $\bg^{-1}\bfid$, the
 square root matrix $\bgamma$ is always real with a 2D trace that never vanishes.
 For the opposite choice it is not guaranteed.   Indeed from Eq.~(\ref{eqn:gammapm}), 
 the solution in our thought example evolves to $\tr{\bgamma}=0$ at $\dot a/\sqrt{-K}=-1$ where an infinite number of solutions exist.
 Beyond this point, it is not even obvious that $\tr{\bgamma}$ and the square root 
 will remain real.    We show in \S \ref{sec:zweibeins}  that on self-accelerating solutions
 that it does for any well-posed initial value problem including this thought example.
 
{If we treat the \stucky\ fields as physical fields and not as mathematical devices, the det$>$0 choice is unnatural:
 it introduces a cusp at the det=0 point.
 This suggests that framing the underlying theory directly in terms of vierbeins, which we shall see are closely related to the \stucky\ Jacobian, may help to resolve this ambiguity.}

\subsection{Vierbeins}
\label{sec:vierbeins}

The metric formulation of massive gravity has an equivalent vierbein formulation 
\cite{Nibbelink:2006sz,Chamseddine:2011mu,Hinterbichler:2012cn,Deffayet:2012zc,Deffayet:2012nr}.  Here
there are two vierbeins, one for each metric.  
We define the respective inverse vierbeins with a Minkowski tetrad metric
\begin{eqnarray}
g_{\mu\nu} & = &\eta_{ab} E^{a}_{\hphantom{a}\mu}  E^{b}_{\hphantom{b}\nu} , \nonumber\\
\Sigma_{\mu\nu} & = &\eta_{ab} L^{a}_{\hphantom{a}\mu}  L^{b}_{\hphantom{b}\nu}  ,
\label{eqn:ivierbeins}
\end{eqnarray}
and the vierbeins as the dual or inverses
\begin{eqnarray}
E^{a}_{\hphantom{a}\mu} e_b^{\hphantom{b}\mu} &=& \delta^a_{\hphantom{a}b}, 
\quad \lu{e}{a}{\mu} \ul{E}{a}{\nu} = \ul{\delta}{\mu}{\nu},
\nonumber\\
L^{a}_{\hphantom{a}\mu} \ell_b^{\hphantom{b}\mu} &=& \delta^a_{\hphantom{a}b},
\quad \lu{\ell}{a}{\mu} \ul{L}{a}{\nu} = \ul{\delta}{\mu}{\nu}.
\end{eqnarray}
Note that if we keep the convention that indices are raised and lowered by the
spacetime metric then Eq.~(\ref{eqn:ivierbeins}) implies 
$E^{a}_{\hphantom{a}\mu} =e^{a}_{\hphantom{a}\mu}$.  We attempt though to avoid 
potential notational ambiguities associated with raising and lowering.   For example when using
matrix notation we define $\bvier \equiv \lu{e}{a}{\mu}$ and explicitly write
the matrix inverse of the fiducial metric as 
 $[\bfid^{-1}]^{\mu\nu} \ne \fid^{\mu\nu}$.   Thus
\begin{eqnarray}
g^{\mu\nu} &=& \eta^{ab}   \lu{e}{a}{\mu}\lu{e}{b}{\nu}  ,\nonumber\\
\left[\bfid^{-1}\right]^{\mu\nu} &=& \eta^{ab}  \lu{\ell}{a}{\mu}  \lu{\ell}{b}{\nu} .
\end{eqnarray}

Massive gravity can be formulated in terms of vierbeins as the dynamical variables, obviating
the need to take explicit square roots of matrices when finding solutions \cite{Nibbelink:2006sz,Hinterbichler:2012cn,Deffayet:2012zc,Hassan:2012wt}.  The relationship between
the solutions of one or the other is established as follows.   If the vierbeins satisfy the symmetry condition
\begin{eqnarray}
e_a^{\hphantom{a}\mu} L_{b\mu} &=& e_b^{\hphantom{b}\mu}  L_{a\mu},
\label{eqn:vsym}
\end{eqnarray}
then there is a real square root $\ul{\gamma}{\mu}{\nu}$,
\begin{equation}
\gamma^{\mu}_{\hphantom{\mu}\alpha} \gamma^\alpha_{\hphantom{\alpha}\nu} = 
g^{\mu\alpha} \Sigma_{\alpha\nu},
\label{eqn:sqrtvierbein}
\end{equation}
given by 
\begin{eqnarray}
\gamma^\mu_{\hphantom{\mu}\nu} &=& e_a^{\hphantom{a}\mu} L^a_{\hphantom{a}\nu} ,
\end{eqnarray}
since
\begin{eqnarray}
\gamma^{\mu}_{\hphantom{\mu}\alpha} \gamma^\alpha_{\hphantom{\alpha}\nu} 
&=& \eta^{ac} \eta^{bd} e_a^{\hphantom{a}\mu} L_{c\alpha}
e_b^{\hphantom{b}\alpha}L_{d\nu}  \nonumber\\
&=& \eta^{ac} \eta^{bd} e_a^{\hphantom{a}\mu} L_{b\alpha} 
e_c^{\hphantom{c}\alpha}L_{d\nu}  \nonumber\\
&=& \eta^{ab} \lu{e}{a}{\mu} \lu{e}{b}{\alpha} \eta_{cd} \ul{L}{c}{\alpha} \ul{L}{d}{\nu} \nonumber\\
&=& g^{\mu\alpha}\Sigma_{\alpha\nu}.
\label{eqn:gammasqrt}
\end{eqnarray}
The square root also satisfies the symmetry condition
\begin{eqnarray}
\fid_{\mu\alpha}\ul{\gamma}{\alpha}{\nu} &=& \ul{L}{a}{\mu} L_{a\alpha} \lu{e}{b}{\alpha}\ul{L}{b}{\nu} \nonumber\\
&=& 
 \ul{L}{a}{\mu} L_{b\alpha} \lu{e}{a}{\alpha}\ul{L}{b}{\nu} \nonumber\\
 &=& \fid_{\nu\alpha}\ul{\gamma}{\alpha}{\mu} .
\label{eqn:fidgammasym}
\end{eqnarray} 

In the other direction, given a real square root that satisfies Eq.~(\ref{eqn:fidgammasym})
the inverse construction
\begin{equation}
\lu{e}{b}{\mu} \equiv \ul{\gamma}{\mu}{\nu} \lu{\ell}{b}{\nu}
\end{equation}
is a vierbein of the spacetime metric
\begin{eqnarray}
\eta^{ab}\lu{e}{a}{\mu} \lu{e}{b}{\nu} &=& 
\ul{\gamma}{\mu}{\alpha} [\bfid^{-1}]^{\alpha\beta} \ul{\gamma}{\nu}{\beta} \nonumber\\
&=& [\bfid^{-1}]^{\mu\gamma} \fid_{\gamma \delta} \ul{\gamma}{\delta}{\alpha}
 [\bfid^{-1}]^{\alpha\beta} \ul{\gamma}{\nu}{\beta} \nonumber\\
 &=&
  [\bfid^{-1}]^{\mu\gamma} \fid_{\alpha \delta} \ul{\gamma}{\delta}{\gamma}
 [\bfid^{-1}]^{\alpha\beta} \ul{\gamma}{\nu}{\beta} \nonumber\\
&=&   [\bfid^{-1}]^{\mu\gamma}  \ul{\gamma}{\delta}{\gamma}\ul{\gamma}{\nu}{\delta} \nonumber\\
&=&  [\bfid^{-1}]^{\mu\gamma}   g^{\nu\delta} \fid_{\delta\gamma} \nonumber\\
&=& g^{\mu\nu}.
\end{eqnarray}
For the block diagonal 2$\times$2 case that we consider below, it is easy to see that the
additional symmetry condition Eq.~(\ref{eqn:fidgammasym}) is an automatic consequence 
of a square root of the form in Eq.~(\ref{eqn:CH}) where $\tr{\bgamma} \ne 0$.  In these cases, each
real square root yields a vierbein satisfying
Eq.~(\ref{eqn:vsym}) by virtue of Eq.~(\ref{eqn:gammasqrt}).  For the special case of the trace singularity
$\tr{\bgamma}=0$, not all of the infinite number of possible real square roots $\bgamma$ satisfy this symmetry though it is 
straightforward to pick one that does.

We can use the definition of the \stucky\ fields from Eq.~(\ref{eqn:stucky})
to relate them to the inverse vierbeins of the fiducial metric:
\begin{equation}
L^a_{\hphantom{a}\mu} = \pm \partial_\mu \phi^a.
\label{eqn:vierstucky}
\end{equation}
Note that in unitary gauge $L^a_{\hphantom{a}\mu} = \pm  \delta^a_{\hphantom{a}\mu}$. 
In this representation,  the only dynamical variables are those
associated with the spacetime vierbeins.   However, this does not automatically solve
the det$[\bfid]=0$ problem.  It is still
the case that the vierbein formulation is not defined here as the inverse vierbein is not
itself invertible.  
On the other hand, one
can demand that the dynamical variables remain continuous across the transition.

\subsection{Zweibeins}
\label{sec:zweibeins}

In our spherically symmetric thought example the relevant part of the vierbein is the 2$\times$2 time-radius block or zweibein.  Strictly speaking, since our \stucky\ basis was originally
defined in Cartesian coordinates in Eq.~(\ref{eqn:fg}), this involves a local rotation of the tetrad basis which
we leave implicit for notational simplicity.    Note that local, spacetime dependent rotations as well as boosts
are allowed as transformations on the vierbeins to bring them into symmetric form
while leaving $\bmin$ invariant (see also \S \ref{sec:imaginary}).

Given each of the possible square root solutions, we can 
construct the associated symmetric vierbein solution from Eq.~(\ref{eqn:sqrtvierbein}) by
inverting the \stucky\ Jacobian.
We can choose the preferred solution as the one where the vierbeins are smooth
across the det=0 singularity.   

In our thought example, the inverse \stucky\ Jacobian 
\begin{align}
\bfvier^{-t} &=
\left(\partial_\mu\phi^a\right)^{-1} \\
&= \frac{1}{x_0}
 \left( 
\begin{array}{cc}
\dfrac{4 - K r^2}{4+K r^2} \dfrac{\sqrt{- K}}{\afd}& -\dfrac{\sqrt{-K}r}{\af}\\
-\dfrac{4} {4+K r^2}\dfrac{-K r}{\afd} & \dfrac{4 -K r^2}{4 \af}\\
\end{array} 
\right) ,\nonumber
\end{align}
and
we start in the expanding phase with 
\begin{equation}
\bgamma_{+} =x_0
\left( 
\begin{array}{cc}
+ \dfrac{\afd}{\sqrt{-K}} & 0 \\
0 & 1 \\
\end{array} 
\right), \quad \afd>0.
\end{equation}
Note that we choose a positive overall sign as the double valued correspondence
in Eq.~(\ref{eqn:vierstucky}) is the same irrelevant 
overall sign ambiguity of the square root.
Using $\bvier = \bfvier^{-t} \bgamma^t$, we obtain
\begin{equation}
 \lu{e}{a}{\mu}
=
\left( 
\begin{array}{cc}
\dfrac{4- K r^2 }{4+K r^2} & -\dfrac{\sqrt{-K}r}{\af} \vphantom{\Bigg[}\\
-\dfrac{4\sqrt{-K} r }{4+K r^2}& \dfrac{4 -K  r^2}{4 \af} \\
\end{array} 
\right), \quad \afd>0.
\end{equation}
On the other side of the turnaround,
\begin{equation}
\lu{e}{a}{\mu} =
 \left( 
\begin{array}{cc}
\pm \dfrac{4- K r^2 }{4+K r^2} & - \dfrac{\sqrt{-K}r}{\af} \vphantom{\Bigg[}\\
\mp \dfrac{4\sqrt{-K} r }{4+K r^2}& \dfrac{4 -K  r^2}{4 \af} \\
\end{array} 
\right), \quad  \afd<0, 
\end{equation}
with $\pm$ corresponding to the zweibein constructed out of $\bgamma_\pm$ 
from Eq.~(\ref{eqn:gammapm}) respectively.
Thus the spacetime zweibein is only continuous if the square root solution is taken to
be 
$\bgamma_+$.   Recall that this is the solution where the determinant changes sign at $\afd=0$.  Thus the positive definite
square root prescription of the metric approach is  not supported  by the vierbein approach.
Furthermore, the zweibein remains continuous even when $\tr{\bgamma_+}=0$ where
there are an infinite number of solutions to the matrix square root.    Beyond this point, the
square root remains real on this solution.  

Note that these results are more general than our particular thought example.   Continuity
of the \stucky\ fields -- or their derivatives, the fiducial vierbeins -- and of the spacetime
vierbeins implies that the matrix square root should not be defined
with a $\mu$ term  in Eq.~(\ref{eqn:mujac}) that enforces an absolute value prescription
for the square root
 as is commonly done in the literature (see also \cite{Fasiello:2012rw}). 
Moreover, a real square root is guaranteed by the \stucky\ equations of motion 
(see also \cite{Nibbelink:2006sz}). 
 For
any isotropic self-accelerating solution,
\begin{equation}
P_1'(x_0) \left( \frac{\det{\bgamma}}{x_0} + x_0 - \tr{\bgamma}\right) =0
\label{eqn:fsoln}
\end{equation}
(see \cite{Gratia:2012wt} Eq.~19 and \cite{Wyman:2012iw} Eq.~16 where $\sqrt{X}=\tr{\bgamma}$, $W=\det\bgamma$).   Except for the special case that $P_1'(x_0)=0$, the reality of the \stucky\ Jacobian and the square root construction of Eq.~(\ref{eqn:CH})
then enforces reality.    This result applies not only to different massive gravity parameters and global cosmological solutions but also to any isotropic
distribution of matter, including black hole solutions; we will consider these in the next section.

The  case $P_1'(x_0)=0$ involves a special choice of the parameters
$\alpha_3$ and $\alpha_4$ \cite{Nieuwenhuizen:2011sq,Berezhiani:2011mt}.  Static black hole solutions which enforce Eq.~(\ref{eqn:fsoln}) by parameter choice rather than dynamics have been constructed to have pathologies 
at the horizon \cite{Nieuwenhuizen:2011sq,Gruzinov:2011mm}.   However doing so leads
to an ill-posed initial value problem: starting from a fiducial metric specified by $f$ and $g$
on an initial surface there is no unique solution for the evolution of $f$.   It is then not surprising that $f$ can be chosen to give a fiducial metric that is sufficiently far from the spacetime metric to be pathological.   We therefore
do not consider such solutions further here.

\section{Singular Spacetime Metric}
\label{sec:deSitter}

Let us also consider cases where the spacetime metric carries singularities while the fiducial metric is well-behaved.   To construct such examples, it 
is simplest to start from unitary gauge where the fiducial metric is Minkowski and look
for spherically symmetric solutions for the spacetime metric.  We begin in \S \ref{sec:imaginary} with a discussion of
why potential new pathologies are related to interchanging the roles of space and time between the
metrics and then show they do not appear in practice for black hole solutions in the
self-accelerating background in \S \ref{sec:SdS}.

\subsection{Space, Time and Imaginary Roots}
\label{sec:imaginary}

It is again instructive to begin by analyzing some general properties of matrix square roots 
and zweibeins.   Since the fiducial metric is Minkowski in unitary gauge
\begin{equation}
\bg^{-1}\bfid = \bvier^t \bmin\, \bvier\, \bmin,
\end{equation}
and given a general zweibein
\begin{equation}
\lu{e}{a}{\mu} =
 \left( 
\begin{array}{cc}
A & B\\
C&D  \\
\end{array} 
\right),
\label{eqn:genzweibein}
\end{equation}
we can determine when pathologies in the square root can occur.

A determinant singularity would occur when $AD = BC$, which is the analogue of the
coordinate type singularity discussed in the previous section but moved from the fiducial
zweibein to the spacetime zweibein.
The zweibein formulation raises the possibility of a different type of problem related
to off-diagonal dominant zweibeins.  Through 
 Eq.~(\ref{eqn:trace}), we can see that all square roots of $\bg^{-1}\bfid$ fail to be
 real if both
 \begin{align}
 (A+D)^2 &< (B+C)^2 , \nonumber\\
 (A-D)^2 &< (B-C)^2 ,
 \end{align}
 and the trace is non-zero.
 Note that these are the same conditions that forbid us from writing the zweibein in Eq.~(\ref{eqn:genzweibein}) as 
 \begin{equation}
 \bvier = \blambda \tilde \bvier,
 \end{equation}
 where 
$\blambda$ is a Lorentz transformation and  $\tilde \bvier$ satisfies the symmetry condition of Eq.~(\ref{eqn:vsym}), $ \bmin\, \tilde \bvier = (\bmin\,\tilde \bvier)^t$.  This generalizes the argument of Ref.~\cite{Deffayet:2012zc} to 
 determinant changing non-orthochronous and improper Lorentz transformations.
 A simple example where these transformations would be required and both a real square
 root and symmetric zweibein exists is given by
 $A=-D$, $B=\epsilon A$ and $C=0$ for any $|\epsilon|<1$.

   In the cases where the real square root ceases to exist, the sense of time and radius between the
 spacetime metric and fiducial metrics are flipped.   For example if $A=D=0$ and
 $B=C+\epsilon=1$ then the 2$\times$2 block of the inverse metric is
  \begin{align}
\bg^{-1} &= \left( 
\begin{array}{cc}
(1-\epsilon)^2  & 0\\
0&-1 \\
\end{array} 
\right), 
\end{align}
and hence
\begin{align}
 \bg^{-1}\bmin &= \left( 
\begin{array}{cc}
-(1-\epsilon)^2   & 0\\
0&-1 \\
\end{array} 
\right) .
\end{align}
Note that strictly speaking for $\epsilon=0$ there are an infinite number of real square roots 
involving the $\tr{\bgamma}=0$ singularity, but for any finite $\epsilon \rightarrow 0$ the roots cease to 
exist (cf.~\cite{Deffayet:2012zc}, Eq.~7.5). These cases quantify what it means for the spacetime metric to be so far from the fiducial
metric as to break the theory.

This is exactly the type of situation that would be encountered in the Schwarzschild 
metric inside the horizon where space flows inevitably into the singularity.    
Given an inverse metric $g^{00} = (p-1)^{-1}$ and $g^{rr} = -(p-1)$ where $p= 2 M/r$, 
the corresponding zweibein is 
\begin{equation}
\lu{e}{a}{\mu} =
 \left( 
\begin{array}{cc}
0  & \sqrt{p-1}\\
\sqrt{\dfrac{1}{p-1}} &0 \\
\end{array} 
\right), \quad p>1,
\label{eqn:schzweibein}
\end{equation}
and a real square root fails to exist.

If this type of solution can be achieved dynamically in massive gravity then it would
represent a more severe type of pathology than the coordinate one encountered in the previous section.   However Eq.~(\ref{eqn:fsoln}) would imply that this pathology is
prevented by the \stucky\ equations of motion on the self-accelerating solution.    The key to this resolution is that
the coordinate system where the fiducial metric
is Minkowski is not necessarily the one for which space and time exchange roles
at the horizon.

\subsection{Schwarzschild de Sitter Case}
\label{sec:SdS}

The simplest example which manifests some of these properties is the static
SdS metric with no vector mode in the background  \cite{Koyama:2011xz,Koyama:2011yg,Koyama:2011wx}.
In unitary gauge, using the notation of Ref.~\cite{Wyman:2012iw},
\begin{align}
 ds^2 &=- \frac{1-p}{x_0^2}df^2 + \frac{1+p}{x_0^2} dg^2 +  \frac{2 p }{x_0^2} df dg
+ \frac{g^2  }{x_0^2} d\Omega^2
\end{align}
where
\begin{equation}
p = \frac{2 M}{g} + \frac{ H^2 g^2}{x_0^2},
\end{equation}
with $M$ as the Schwarzschild mass term and $H^2 = P_0(x_0) m^2/6$ by virtue of the 
self-accelerating effective cosmological constant (see Eq.~\ref{gsol}).
Recall here that unitary time $\tilde t=f$ and radial coordinate $\tilde r=g$.    Crucially,
this chart differs from the standard SdS chart in that none of the metric elements
diverge at the Schwarzschild horizon but $\tilde g_{00}$ can still vanish.

In this case the fiducial metric is Minkowski and therefore 
\begin{equation}
\tilde \bg^{-1} \bfid
=x_0^2 \left(
      \begin{array}{cccc}
        1+p & p & 0 & 0 \\
        -p & 1-p & 0 & 0\\
        0 & 0 & 1& 0\\
        0 & 0 & 0 &1
      \end{array} \right).
\end{equation}
The determinant of this matrix is $x_0^8$ and so in spite of the spacetime singularities,
this leads to a well-defined square root throughout the static spacetime.  Likewise
spacetime scalars associated with $\bg^{-1}\bfid$ remain finite except at the origin.  For the
relevant 2$\times$2 submatrix, the det$>$0 square root is
\begin{equation}
\tilde \bgamma = x_0
\left( 
\begin{array}{cc}
1+p/2 & p/2 \\
-p/2 & 1-p/2 \\
\end{array} 
\right),
\end{equation}
and the zweibein is
\begin{equation}
\lu{\tilde e}{a}{\mu} =
x_0 \left( 
\begin{array}{cc}
1+p/2 & -p/2 \\
p/2 & 1-p/2 \\
\end{array} 
\right).
\end{equation}
Neither of these quantities exhibit problems at the Schwarzschild horizon.  

On the other hand,
the static SdS spacetime encounters another coordinate singularity at
the  cosmological
horizon
 of the observer at the origin.    Here we take the Schwarzschild mass $M=0$ and transform
to the expanding de Sitter coordinates in isotropic gauge
\begin{equation}
ds^2 = -dt^2 + \af^2(t) (dr^2 + r^2 d\Omega^2),
\end{equation}
with the relations
\begin{align}
f(r,t) &= \frac{x_0}{ H }\left[ H t -  z +  \ln \bigg| \frac{1+z} { 1 - z^2}\bigg|
\right],
\nonumber\\
g(r,t) &= x_0 \af(t) r ,
\end{align}
where $z= \af H r$.   Here
 we have continued the solution beyond the coordinate singularity of the SdS chart 
at the cosmological horizon $z=1$.   Note that there is a missing absolute value in Eq.~(40) of Ref.~\cite{Wyman:2012iw},
 where it
was also shown that these \stucky\ fields solve the massive gravity equations for the
full range $0 < r < \infty$.

{We can calculate the vierbeins for this case in two ways: first, by direct calculation in isotropic coordinates starting from the det$>$0 solution at $z\ll 1$, since det($\bg^{-1} \bfid)=x_0^8$ across the full spacetime; or more simply by noticing that the spacetime index of the unitary gauge vierbein transforms as a spacetime tensor.}    Furthermore, the symmetry condition in Eq.~(\ref{eqn:vsym}) is 
spacetime coordinate invariant.
Either procedure yields
\begin{equation}
\lu{ e}{a}{\mu} =\frac{1}{2-2 z}
\left( 
\begin{array}{cc}
2 + z(z -2)  &  \dfrac{z(z -2)}{a} \vphantom{\Bigg[}  \\
 {z( z-2)}  & \dfrac{2+z(z-2)}{a} \\
\end{array} 
\right),
\end{equation}
where $z=a H r$.  The spacetime zweibein has a pole in its amplitude at the cosmological
horizon since
the tetrad basis is aligned with null directions in the spacetime as a manifestation
of the coordinate singularity of the static 
SdS chart.

\section{Discussion}
\label{sec:discussion}

We have shown that massive gravity with a non-dynamical fiducial metric allows
the spacetime metric to evolve far enough away from the fiducial metric so as to
encounter a determinant singularity from non-singular initial conditions.   At this singularity
both the metric and vierbein formulation of the theory become ill-defined.    
In the metric version, a single unitary chart of the spacetime fails to cover the spacetime.
At the singularity, there are multiple solutions of the metric square root and correspondingly
multiple unitary charts on the other side of the singularity.  Likewise, in the vierbein formulation, the fiducial vierbeins become non-invertible at this point.   

On the other hand both the vierbein and \stucky\ formulations suggest that that this
singularity is coordinate related rather than a more fundamental problem.   Using the
vierbein formulation we have shown that the dynamical object, the spacetime vierbein,
remains continuous across the singularity for one choice of how to continue the
unitary chart.  For this choice, the determinant is allowed to switch signs, in contradiction
to a common convention in the literature that maintains the square roots should
remain positive definite \cite{D'Amico:2011jj,Gumrukcuoglu:2011zh}.   

A more fundamental problem would occur if the spacetime metric could evolve to a point
where real square roots or symmetry between the spacetime and fiducial vierbeins
cease to exist.   We have shown that these points occur when the sense of
space and time in the spacetime metric flip, e.g.\ at the horizon of a black hole.
On the other hand we have shown that on the self-accelerating branch, the theory
can accommodate any symmetric distribution of matter, even that leading to the formation
of a black hole.    The resolution of this apparent paradox is that the coordinates
for which the fiducial metric is Minkowski are not the standard Schwarzschild coordinates.
While these examples do not resolve all the potential pathologies of massive gravity
(see e.g.~\cite{Tasinato:2012ze,DeFelice:2012mx,Koyama:2011wx,Burrage:2011cr,Deser:2012qx,DeFelice:2013awa,Izumi:2013poa})
they do indicate that the theory is more robust than might naively be expected.

\smallskip{\em Acknowledgments.--}  We thank D. Holz, A. Tolley, R. Wald, L. Wang for helpful discussions. 
PG was supported by the National Research Fund Luxembourg through grant BFR08-024. WH was supported by Kavli Institute for Cosmological Physics at the University of Chicago through grants NSF PHY-0114422 and NSF PHY-0551142  and an endowment from the Kavli Foundation and its founder Fred Kavli and by the David and Lucile Packard Foundation. MW was supported, and WH was additionally supported, by U.S.~Dept.\ of Energy contract DE-FG02-90ER-40560.

\bibliography{zerodet}

\begin{thebibliography}{34}%
\makeatletter
\providecommand \@ifxundefined [1]{%
 \@ifx{#1\undefined}
}%
\providecommand \@ifnum [1]{%
 \ifnum #1\expandafter \@firstoftwo
 \else \expandafter \@secondoftwo
 \fi
}%
\providecommand \@ifx [1]{%
 \ifx #1\expandafter \@firstoftwo
 \else \expandafter \@secondoftwo
 \fi
}%
\providecommand \natexlab [1]{#1}%
\providecommand \enquote  [1]{``#1''}%
\providecommand \bibnamefont  [1]{#1}%
\providecommand \bibfnamefont [1]{#1}%
\providecommand \citenamefont [1]{#1}%
\providecommand \href@noop [0]{\@secondoftwo}%
\providecommand \href [0]{\begingroup \@sanitize@url \@href}%
\providecommand \@href[1]{\@@startlink{#1}\@@href}%
\providecommand \@@href[1]{\endgroup#1\@@endlink}%
\providecommand \@sanitize@url [0]{\catcode `\\12\catcode `\$12\catcode
  `\&12\catcode `\#12\catcode `\^12\catcode `\_12\catcode `\%12\relax}%
\providecommand \@@startlink[1]{}%
\providecommand \@@endlink[0]{}%
\providecommand \url  [0]{\begingroup\@sanitize@url \@url }%
\providecommand \@url [1]{\endgroup\@href {#1}{\urlprefix }}%
\providecommand \urlprefix  [0]{URL }%
\providecommand \Eprint [0]{\href }%
\providecommand \doibase [0]{http://dx.doi.org/}%
\providecommand \selectlanguage [0]{\@gobble}%
\providecommand \bibinfo  [0]{\@secondoftwo}%
\providecommand \bibfield  [0]{\@secondoftwo}%
\providecommand \translation [1]{[#1]}%
\providecommand \BibitemOpen [0]{}%
\providecommand \bibitemStop [0]{}%
\providecommand \bibitemNoStop [0]{.\EOS\space}%
\providecommand \EOS [0]{\spacefactor3000\relax}%
\providecommand \BibitemShut  [1]{\csname bibitem#1\endcsname}%
\let\auto@bib@innerbib\@empty
\bibitem [{\citenamefont {Fierz}\ and\ \citenamefont
  {Pauli}(1939)}]{Fierz:1939ix}%
  \BibitemOpen
  \bibfield  {author} {\bibinfo {author} {\bibfnamefont {M.}~\bibnamefont
  {Fierz}}\ and\ \bibinfo {author} {\bibfnamefont {W.}~\bibnamefont {Pauli}},\
  }\href@noop {} {\bibfield  {journal} {\bibinfo  {journal}
  {Proc.Roy.Soc.Lond.}\ }\textbf {\bibinfo {volume} {A173}},\ \bibinfo {pages}
  {211} (\bibinfo {year} {1939})}\BibitemShut {NoStop}%
\bibitem [{\citenamefont {Boulware}\ and\ \citenamefont
  {Deser}(1972)}]{Boulware:1972zf}%
  \BibitemOpen
  \bibfield  {author} {\bibinfo {author} {\bibfnamefont {D.}~\bibnamefont
  {Boulware}}\ and\ \bibinfo {author} {\bibfnamefont {S.}~\bibnamefont
  {Deser}},\ }\href {\doibase 10.1016/0370-2693(72)90418-2} {\bibfield
  {journal} {\bibinfo  {journal} {Phys.Lett.}\ }\textbf {\bibinfo {volume}
  {B40}},\ \bibinfo {pages} {227} (\bibinfo {year} {1972})}\BibitemShut
  {NoStop}%
\bibitem [{\citenamefont {de~Rham}\ and\ \citenamefont
  {Gabadadze}(2010)}]{deRham:2010ik}%
  \BibitemOpen
  \bibfield  {author} {\bibinfo {author} {\bibfnamefont {C.}~\bibnamefont
  {de~Rham}}\ and\ \bibinfo {author} {\bibfnamefont {G.}~\bibnamefont
  {Gabadadze}},\ }\href {\doibase 10.1103/PhysRevD.82.044020} {\bibfield
  {journal} {\bibinfo  {journal} {Phys.Rev.}\ }\textbf {\bibinfo {volume}
  {D82}},\ \bibinfo {pages} {044020} (\bibinfo {year} {2010})},\ \Eprint
  {http://arxiv.org/abs/1007.0443} {arXiv:1007.0443 [hep-th]} \BibitemShut
  {NoStop}%
\bibitem [{\citenamefont {de~Rham}\ \emph {et~al.}(2011)\citenamefont
  {de~Rham}, \citenamefont {Gabadadze},\ and\ \citenamefont
  {Tolley}}]{deRham:2010kj}%
  \BibitemOpen
  \bibfield  {author} {\bibinfo {author} {\bibfnamefont {C.}~\bibnamefont
  {de~Rham}}, \bibinfo {author} {\bibfnamefont {G.}~\bibnamefont {Gabadadze}},
  \ and\ \bibinfo {author} {\bibfnamefont {A.~J.}\ \bibnamefont {Tolley}},\
  }\href {\doibase 10.1103/PhysRevLett.106.231101} {\bibfield  {journal}
  {\bibinfo  {journal} {Phys.Rev.Lett.}\ }\textbf {\bibinfo {volume} {106}},\
  \bibinfo {pages} {231101} (\bibinfo {year} {2011})},\ \Eprint
  {http://arxiv.org/abs/1011.1232} {arXiv:1011.1232 [hep-th]} \BibitemShut
  {NoStop}%
\bibitem [{\citenamefont {Hassan}\ and\ \citenamefont
  {Rosen}(2012)}]{Hassan:2011hr}%
  \BibitemOpen
  \bibfield  {author} {\bibinfo {author} {\bibfnamefont {S.}~\bibnamefont
  {Hassan}}\ and\ \bibinfo {author} {\bibfnamefont {R.~A.}\ \bibnamefont
  {Rosen}},\ }\href {\doibase 10.1103/PhysRevLett.108.041101} {\bibfield
  {journal} {\bibinfo  {journal} {Phys.Rev.Lett.}\ }\textbf {\bibinfo {volume}
  {108}},\ \bibinfo {pages} {041101} (\bibinfo {year} {2012})},\ \Eprint
  {http://arxiv.org/abs/1106.3344} {arXiv:1106.3344 [hep-th]} \BibitemShut
  {NoStop}%
\bibitem [{\citenamefont {Deffayet}\ and\ \citenamefont
  {Jacobson}(2012)}]{Deffayet:2011rh}%
  \BibitemOpen
  \bibfield  {author} {\bibinfo {author} {\bibfnamefont {C.}~\bibnamefont
  {Deffayet}}\ and\ \bibinfo {author} {\bibfnamefont {T.}~\bibnamefont
  {Jacobson}},\ }\href {\doibase 10.1088/0264-9381/29/6/065009} {\bibfield
  {journal} {\bibinfo  {journal} {Class.Quant.Grav.}\ }\textbf {\bibinfo
  {volume} {29}},\ \bibinfo {pages} {065009} (\bibinfo {year} {2012})},\
  \Eprint {http://arxiv.org/abs/1107.4978} {arXiv:1107.4978 [gr-qc]}
  \BibitemShut {NoStop}%
\bibitem [{\citenamefont {Volkov}(2013)}]{Volkov:2013roa}%
  \BibitemOpen
  \bibfield  {author} {\bibinfo {author} {\bibfnamefont {M.~S.}\ \bibnamefont
  {Volkov}},\ }\href@noop {} {\  (\bibinfo {year} {2013})},\ \Eprint
  {http://arxiv.org/abs/1304.0238} {arXiv:1304.0238 [hep-th]} \BibitemShut
  {NoStop}%
\bibitem [{\citenamefont {D'Amico}\ \emph {et~al.}(2011)\citenamefont
  {D'Amico}, \citenamefont {de~Rham}, \citenamefont {Dubovsky}, \citenamefont
  {Gabadadze}, \citenamefont {Pirtskhalava} \emph {et~al.}}]{D'Amico:2011jj}%
  \BibitemOpen
  \bibfield  {author} {\bibinfo {author} {\bibfnamefont {G.}~\bibnamefont
  {D'Amico}}, \bibinfo {author} {\bibfnamefont {C.}~\bibnamefont {de~Rham}},
  \bibinfo {author} {\bibfnamefont {S.}~\bibnamefont {Dubovsky}}, \bibinfo
  {author} {\bibfnamefont {G.}~\bibnamefont {Gabadadze}}, \bibinfo {author}
  {\bibfnamefont {D.}~\bibnamefont {Pirtskhalava}},  \emph {et~al.},\ }\href
  {\doibase 10.1103/PhysRevD.84.124046} {\bibfield  {journal} {\bibinfo
  {journal} {Phys.Rev.}\ }\textbf {\bibinfo {volume} {D84}},\ \bibinfo {pages}
  {124046} (\bibinfo {year} {2011})},\ \Eprint {http://arxiv.org/abs/1108.5231}
  {arXiv:1108.5231 [hep-th]} \BibitemShut {NoStop}%
\bibitem [{\citenamefont {Gumrukcuoglu}\ \emph {et~al.}(2011)\citenamefont
  {Gumrukcuoglu}, \citenamefont {Lin},\ and\ \citenamefont
  {Mukohyama}}]{Gumrukcuoglu:2011ew}%
  \BibitemOpen
  \bibfield  {author} {\bibinfo {author} {\bibfnamefont {A.~E.}\ \bibnamefont
  {Gumrukcuoglu}}, \bibinfo {author} {\bibfnamefont {C.}~\bibnamefont {Lin}}, \
  and\ \bibinfo {author} {\bibfnamefont {S.}~\bibnamefont {Mukohyama}},\
  }\href@noop {} {\bibfield  {journal} {\bibinfo  {journal} {JCAP}\ }\textbf
  {\bibinfo {volume} {1111}},\ \bibinfo {pages} {030} (\bibinfo {year}
  {2011})},\ \Eprint {http://arxiv.org/abs/1109.3845} {arXiv:1109.3845
  [hep-th]} \BibitemShut {NoStop}%
\bibitem [{\citenamefont {Koyama}\ \emph
  {et~al.}(2011{\natexlab{a}})\citenamefont {Koyama}, \citenamefont {Niz},\
  and\ \citenamefont {Tasinato}}]{Koyama:2011xz}%
  \BibitemOpen
  \bibfield  {author} {\bibinfo {author} {\bibfnamefont {K.}~\bibnamefont
  {Koyama}}, \bibinfo {author} {\bibfnamefont {G.}~\bibnamefont {Niz}}, \ and\
  \bibinfo {author} {\bibfnamefont {G.}~\bibnamefont {Tasinato}},\ }\href@noop
  {} {\bibfield  {journal} {\bibinfo  {journal} {Phys.Rev.Lett.}\ }\textbf
  {\bibinfo {volume} {107}},\ \bibinfo {pages} {131101} (\bibinfo {year}
  {2011}{\natexlab{a}})},\ \Eprint {http://arxiv.org/abs/1103.4708}
  {arXiv:1103.4708 [hep-th]} \BibitemShut {NoStop}%
\bibitem [{\citenamefont {Koyama}\ \emph
  {et~al.}(2011{\natexlab{b}})\citenamefont {Koyama}, \citenamefont {Niz},\
  and\ \citenamefont {Tasinato}}]{Koyama:2011yg}%
  \BibitemOpen
  \bibfield  {author} {\bibinfo {author} {\bibfnamefont {K.}~\bibnamefont
  {Koyama}}, \bibinfo {author} {\bibfnamefont {G.}~\bibnamefont {Niz}}, \ and\
  \bibinfo {author} {\bibfnamefont {G.}~\bibnamefont {Tasinato}},\ }\href@noop
  {} {\bibfield  {journal} {\bibinfo  {journal} {Phys.Rev.}\ }\textbf {\bibinfo
  {volume} {D84}},\ \bibinfo {pages} {064033} (\bibinfo {year}
  {2011}{\natexlab{b}})},\ \Eprint {http://arxiv.org/abs/1104.2143}
  {arXiv:1104.2143 [hep-th]} \BibitemShut {NoStop}%
\bibitem [{\citenamefont {Nieuwenhuizen}(2011)}]{Nieuwenhuizen:2011sq}%
  \BibitemOpen
  \bibfield  {author} {\bibinfo {author} {\bibfnamefont {T.}~\bibnamefont
  {Nieuwenhuizen}},\ }\href {\doibase 10.1103/PhysRevD.84.024038} {\bibfield
  {journal} {\bibinfo  {journal} {Phys.Rev.}\ }\textbf {\bibinfo {volume}
  {D84}},\ \bibinfo {pages} {024038} (\bibinfo {year} {2011})},\ \Eprint
  {http://arxiv.org/abs/1103.5912} {arXiv:1103.5912 [gr-qc]} \BibitemShut
  {NoStop}%
\bibitem [{\citenamefont {Berezhiani}\ \emph {et~al.}(2012)\citenamefont
  {Berezhiani}, \citenamefont {Chkareuli}, \citenamefont {de~Rham},
  \citenamefont {Gabadadze},\ and\ \citenamefont {Tolley}}]{Berezhiani:2011mt}%
  \BibitemOpen
  \bibfield  {author} {\bibinfo {author} {\bibfnamefont {L.}~\bibnamefont
  {Berezhiani}}, \bibinfo {author} {\bibfnamefont {G.}~\bibnamefont
  {Chkareuli}}, \bibinfo {author} {\bibfnamefont {C.}~\bibnamefont {de~Rham}},
  \bibinfo {author} {\bibfnamefont {G.}~\bibnamefont {Gabadadze}}, \ and\
  \bibinfo {author} {\bibfnamefont {A.}~\bibnamefont {Tolley}},\ }\href
  {\doibase 10.1103/PhysRevD.85.044024} {\bibfield  {journal} {\bibinfo
  {journal} {Phys.Rev.}\ }\textbf {\bibinfo {volume} {D85}},\ \bibinfo {pages}
  {044024} (\bibinfo {year} {2012})},\ \Eprint {http://arxiv.org/abs/1111.3613}
  {arXiv:1111.3613 [hep-th]} \BibitemShut {NoStop}%
\bibitem [{\citenamefont {Gratia}\ \emph {et~al.}(2012)\citenamefont {Gratia},
  \citenamefont {Hu},\ and\ \citenamefont {Wyman}}]{Gratia:2012wt}%
  \BibitemOpen
  \bibfield  {author} {\bibinfo {author} {\bibfnamefont {P.}~\bibnamefont
  {Gratia}}, \bibinfo {author} {\bibfnamefont {W.}~\bibnamefont {Hu}}, \ and\
  \bibinfo {author} {\bibfnamefont {M.}~\bibnamefont {Wyman}},\ }\href
  {\doibase 10.1103/PhysRevD.86.061504} {\bibfield  {journal} {\bibinfo
  {journal} {Phys.Rev.}\ }\textbf {\bibinfo {volume} {D86}},\ \bibinfo {pages}
  {061504} (\bibinfo {year} {2012})},\ \Eprint {http://arxiv.org/abs/1205.4241}
  {arXiv:1205.4241 [hep-th]} \BibitemShut {NoStop}%
\bibitem [{\citenamefont {Kobayashi}\ \emph {et~al.}(2012)\citenamefont
  {Kobayashi}, \citenamefont {Siino}, \citenamefont {Yamaguchi},\ and\
  \citenamefont {Yoshida}}]{Kobayashi:2012fz}%
  \BibitemOpen
  \bibfield  {author} {\bibinfo {author} {\bibfnamefont {T.}~\bibnamefont
  {Kobayashi}}, \bibinfo {author} {\bibfnamefont {M.}~\bibnamefont {Siino}},
  \bibinfo {author} {\bibfnamefont {M.}~\bibnamefont {Yamaguchi}}, \ and\
  \bibinfo {author} {\bibfnamefont {D.}~\bibnamefont {Yoshida}},\ }\href
  {\doibase 10.1103/PhysRevD.86.061505} {\bibfield  {journal} {\bibinfo
  {journal} {Phys.Rev.}\ }\textbf {\bibinfo {volume} {D86}},\ \bibinfo {pages}
  {061505} (\bibinfo {year} {2012})},\ \Eprint {http://arxiv.org/abs/1205.4938}
  {arXiv:1205.4938 [hep-th]} \BibitemShut {NoStop}%
\bibitem [{\citenamefont {Volkov}(2012{\natexlab{a}})}]{Volkov:2012cf}%
  \BibitemOpen
  \bibfield  {author} {\bibinfo {author} {\bibfnamefont {M.~S.}\ \bibnamefont
  {Volkov}},\ }\href {\doibase 10.1103/PhysRevD.86.061502} {\bibfield
  {journal} {\bibinfo  {journal} {Phys.Rev.}\ }\textbf {\bibinfo {volume}
  {D86}},\ \bibinfo {pages} {061502} (\bibinfo {year} {2012}{\natexlab{a}})},\
  \Eprint {http://arxiv.org/abs/1205.5713} {arXiv:1205.5713 [hep-th]}
  \BibitemShut {NoStop}%
\bibitem [{\citenamefont {Volkov}(2012{\natexlab{b}})}]{Volkov:2012zb}%
  \BibitemOpen
  \bibfield  {author} {\bibinfo {author} {\bibfnamefont {M.~S.}\ \bibnamefont
  {Volkov}},\ }\href {\doibase 10.1103/PhysRevD.86.104022} {\bibfield
  {journal} {\bibinfo  {journal} {Phys.Rev.}\ }\textbf {\bibinfo {volume}
  {D86}},\ \bibinfo {pages} {104022} (\bibinfo {year} {2012}{\natexlab{b}})},\
  \Eprint {http://arxiv.org/abs/1207.3723} {arXiv:1207.3723 [hep-th]}
  \BibitemShut {NoStop}%
\bibitem [{\citenamefont {Wyman}\ \emph {et~al.}(2013)\citenamefont {Wyman},
  \citenamefont {Hu},\ and\ \citenamefont {Gratia}}]{Wyman:2012iw}%
  \BibitemOpen
  \bibfield  {author} {\bibinfo {author} {\bibfnamefont {M.}~\bibnamefont
  {Wyman}}, \bibinfo {author} {\bibfnamefont {W.}~\bibnamefont {Hu}}, \ and\
  \bibinfo {author} {\bibfnamefont {P.}~\bibnamefont {Gratia}},\ }\href
  {\doibase 10.1103/PhysRevD.87.084046} {\bibfield  {journal} {\bibinfo
  {journal} {Phys. Rev. D}\ }\textbf {\bibinfo {volume} {87}},\ \bibinfo
  {pages} {084046} (\bibinfo {year} {2013})},\ \Eprint
  {http://arxiv.org/abs/1211.4576} {arXiv:1211.4576 [hep-th]} \BibitemShut
  {NoStop}%
\bibitem [{\citenamefont {Gumrukcuoglu}\ \emph {et~al.}(2012)\citenamefont
  {Gumrukcuoglu}, \citenamefont {Lin},\ and\ \citenamefont
  {Mukohyama}}]{Gumrukcuoglu:2011zh}%
  \BibitemOpen
  \bibfield  {author} {\bibinfo {author} {\bibfnamefont {A.~E.}\ \bibnamefont
  {Gumrukcuoglu}}, \bibinfo {author} {\bibfnamefont {C.}~\bibnamefont {Lin}}, \
  and\ \bibinfo {author} {\bibfnamefont {S.}~\bibnamefont {Mukohyama}},\
  }\href@noop {} {\bibfield  {journal} {\bibinfo  {journal} {JCAP}\ }\textbf
  {\bibinfo {volume} {1203}},\ \bibinfo {pages} {006} (\bibinfo {year}
  {2012})},\ \Eprint {http://arxiv.org/abs/1111.4107} {arXiv:1111.4107
  [hep-th]} \BibitemShut {NoStop}%
\bibitem [{\citenamefont {Nibbelink~Groot}\ \emph {et~al.}(2007)\citenamefont
  {Nibbelink~Groot}, \citenamefont {Peloso},\ and\ \citenamefont
  {Sexton}}]{Nibbelink:2006sz}%
  \BibitemOpen
  \bibfield  {author} {\bibinfo {author} {\bibfnamefont {S.}~\bibnamefont
  {Nibbelink~Groot}}, \bibinfo {author} {\bibfnamefont {M.}~\bibnamefont
  {Peloso}}, \ and\ \bibinfo {author} {\bibfnamefont {M.}~\bibnamefont
  {Sexton}},\ }\href {\doibase 10.1140/epjc/s10052-007-0311-x} {\bibfield
  {journal} {\bibinfo  {journal} {Eur.Phys.J.}\ }\textbf {\bibinfo {volume}
  {C51}},\ \bibinfo {pages} {741} (\bibinfo {year} {2007})},\ \Eprint
  {http://arxiv.org/abs/hep-th/0610169} {arXiv:hep-th/0610169 [hep-th]}
  \BibitemShut {NoStop}%
\bibitem [{\citenamefont {Chamseddine}\ and\ \citenamefont
  {Mukhanov}(2011)}]{Chamseddine:2011mu}%
  \BibitemOpen
  \bibfield  {author} {\bibinfo {author} {\bibfnamefont {A.~H.}\ \bibnamefont
  {Chamseddine}}\ and\ \bibinfo {author} {\bibfnamefont {V.}~\bibnamefont
  {Mukhanov}},\ }\href {\doibase 10.1007/JHEP08(2011)091} {\bibfield  {journal}
  {\bibinfo  {journal} {JHEP}\ }\textbf {\bibinfo {volume} {1108}},\ \bibinfo
  {pages} {091} (\bibinfo {year} {2011})},\ \Eprint
  {http://arxiv.org/abs/1106.5868} {arXiv:1106.5868 [hep-th]} \BibitemShut
  {NoStop}%
\bibitem [{\citenamefont {Hinterbichler}\ and\ \citenamefont
  {Rosen}(2012)}]{Hinterbichler:2012cn}%
  \BibitemOpen
  \bibfield  {author} {\bibinfo {author} {\bibfnamefont {K.}~\bibnamefont
  {Hinterbichler}}\ and\ \bibinfo {author} {\bibfnamefont {R.~A.}\ \bibnamefont
  {Rosen}},\ }\href {\doibase 10.1007/JHEP07(2012)047} {\bibfield  {journal}
  {\bibinfo  {journal} {JHEP}\ }\textbf {\bibinfo {volume} {1207}},\ \bibinfo
  {pages} {047} (\bibinfo {year} {2012})},\ \Eprint
  {http://arxiv.org/abs/1203.5783} {arXiv:1203.5783 [hep-th]} \BibitemShut
  {NoStop}%
\bibitem [{\citenamefont {Deffayet}\ \emph
  {et~al.}(2013{\natexlab{a}})\citenamefont {Deffayet}, \citenamefont
  {Mourad},\ and\ \citenamefont {Zahariade}}]{Deffayet:2012zc}%
  \BibitemOpen
  \bibfield  {author} {\bibinfo {author} {\bibfnamefont {C.}~\bibnamefont
  {Deffayet}}, \bibinfo {author} {\bibfnamefont {J.}~\bibnamefont {Mourad}}, \
  and\ \bibinfo {author} {\bibfnamefont {G.}~\bibnamefont {Zahariade}},\ }\href
  {\doibase 10.1007/JHEP03(2013)086} {\bibfield  {journal} {\bibinfo  {journal}
  {JHEP}\ }\textbf {\bibinfo {volume} {1303}},\ \bibinfo {pages} {086}
  (\bibinfo {year} {2013}{\natexlab{a}})},\ \Eprint
  {http://arxiv.org/abs/1208.4493} {arXiv:1208.4493 [gr-qc]} \BibitemShut
  {NoStop}%
\bibitem [{\citenamefont {Deffayet}\ \emph
  {et~al.}(2013{\natexlab{b}})\citenamefont {Deffayet}, \citenamefont
  {Mourad},\ and\ \citenamefont {Zahariade}}]{Deffayet:2012nr}%
  \BibitemOpen
  \bibfield  {author} {\bibinfo {author} {\bibfnamefont {C.}~\bibnamefont
  {Deffayet}}, \bibinfo {author} {\bibfnamefont {J.}~\bibnamefont {Mourad}}, \
  and\ \bibinfo {author} {\bibfnamefont {G.}~\bibnamefont {Zahariade}},\ }\href
  {\doibase 10.1088/1475-7516/2013/01/032} {\bibfield  {journal} {\bibinfo
  {journal} {JCAP}\ }\textbf {\bibinfo {volume} {1301}},\ \bibinfo {pages}
  {032} (\bibinfo {year} {2013}{\natexlab{b}})},\ \Eprint
  {http://arxiv.org/abs/1207.6338} {arXiv:1207.6338 [hep-th]} \BibitemShut
  {NoStop}%
\bibitem [{\citenamefont {Hassan}\ \emph {et~al.}(2012)\citenamefont {Hassan},
  \citenamefont {Schmidt-May},\ and\ \citenamefont {von
  Strauss}}]{Hassan:2012wt}%
  \BibitemOpen
  \bibfield  {author} {\bibinfo {author} {\bibfnamefont {S.}~\bibnamefont
  {Hassan}}, \bibinfo {author} {\bibfnamefont {A.}~\bibnamefont {Schmidt-May}},
  \ and\ \bibinfo {author} {\bibfnamefont {M.}~\bibnamefont {von Strauss}},\
  }\href@noop {} {\  (\bibinfo {year} {2012})},\ \Eprint
  {http://arxiv.org/abs/1204.5202} {arXiv:1204.5202 [hep-th]} \BibitemShut
  {NoStop}%
\bibitem [{\citenamefont {Fasiello}\ and\ \citenamefont
  {Tolley}(2012)}]{Fasiello:2012rw}%
  \BibitemOpen
  \bibfield  {author} {\bibinfo {author} {\bibfnamefont {M.}~\bibnamefont
  {Fasiello}}\ and\ \bibinfo {author} {\bibfnamefont {A.~J.}\ \bibnamefont
  {Tolley}},\ }\href {\doibase 10.1088/1475-7516/2012/11/035} {\bibfield
  {journal} {\bibinfo  {journal} {JCAP}\ }\textbf {\bibinfo {volume} {1211}},\
  \bibinfo {pages} {035} (\bibinfo {year} {2012})},\ \Eprint
  {http://arxiv.org/abs/1206.3852} {arXiv:1206.3852 [hep-th]} \BibitemShut
  {NoStop}%
\bibitem [{\citenamefont {Gruzinov}\ and\ \citenamefont
  {Mirbabayi}(2011)}]{Gruzinov:2011mm}%
  \BibitemOpen
  \bibfield  {author} {\bibinfo {author} {\bibfnamefont {A.}~\bibnamefont
  {Gruzinov}}\ and\ \bibinfo {author} {\bibfnamefont {M.}~\bibnamefont
  {Mirbabayi}},\ }\href {\doibase 10.1103/PhysRevD.84.124019} {\bibfield
  {journal} {\bibinfo  {journal} {Phys.Rev.}\ }\textbf {\bibinfo {volume}
  {D84}},\ \bibinfo {pages} {124019} (\bibinfo {year} {2011})},\ \Eprint
  {http://arxiv.org/abs/1106.2551} {arXiv:1106.2551 [hep-th]} \BibitemShut
  {NoStop}%
\bibitem [{\citenamefont {Koyama}\ \emph
  {et~al.}(2011{\natexlab{c}})\citenamefont {Koyama}, \citenamefont {Niz},\
  and\ \citenamefont {Tasinato}}]{Koyama:2011wx}%
  \BibitemOpen
  \bibfield  {author} {\bibinfo {author} {\bibfnamefont {K.}~\bibnamefont
  {Koyama}}, \bibinfo {author} {\bibfnamefont {G.}~\bibnamefont {Niz}}, \ and\
  \bibinfo {author} {\bibfnamefont {G.}~\bibnamefont {Tasinato}},\ }\href
  {\doibase 10.1007/JHEP12(2011)065} {\bibfield  {journal} {\bibinfo  {journal}
  {JHEP}\ }\textbf {\bibinfo {volume} {1112}},\ \bibinfo {pages} {065}
  (\bibinfo {year} {2011}{\natexlab{c}})},\ \Eprint
  {http://arxiv.org/abs/1110.2618} {arXiv:1110.2618 [hep-th]} \BibitemShut
  {NoStop}%
\bibitem [{\citenamefont {Tasinato}\ \emph {et~al.}(2012)\citenamefont
  {Tasinato}, \citenamefont {Koyama},\ and\ \citenamefont
  {Niz}}]{Tasinato:2012ze}%
  \BibitemOpen
  \bibfield  {author} {\bibinfo {author} {\bibfnamefont {G.}~\bibnamefont
  {Tasinato}}, \bibinfo {author} {\bibfnamefont {K.}~\bibnamefont {Koyama}}, \
  and\ \bibinfo {author} {\bibfnamefont {G.}~\bibnamefont {Niz}},\ }\href@noop
  {} {\  (\bibinfo {year} {2012})},\ \Eprint {http://arxiv.org/abs/1210.3627}
  {arXiv:1210.3627 [hep-th]} \BibitemShut {NoStop}%
\bibitem [{\citenamefont {De~Felice}\ \emph {et~al.}(2012)\citenamefont
  {De~Felice}, \citenamefont {Gumrukcuoglu},\ and\ \citenamefont
  {Mukohyama}}]{DeFelice:2012mx}%
  \BibitemOpen
  \bibfield  {author} {\bibinfo {author} {\bibfnamefont {A.}~\bibnamefont
  {De~Felice}}, \bibinfo {author} {\bibfnamefont {A.~E.}\ \bibnamefont
  {Gumrukcuoglu}}, \ and\ \bibinfo {author} {\bibfnamefont {S.}~\bibnamefont
  {Mukohyama}},\ }\href@noop {} {\bibfield  {journal} {\bibinfo  {journal}
  {Phys.Rev.Lett.}\ }\textbf {\bibinfo {volume} {109}},\ \bibinfo {pages}
  {171101} (\bibinfo {year} {2012})},\ \Eprint {http://arxiv.org/abs/1206.2080}
  {arXiv:1206.2080 [hep-th]} \BibitemShut {NoStop}%
\bibitem [{\citenamefont {Burrage}\ \emph {et~al.}(2012)\citenamefont
  {Burrage}, \citenamefont {de~Rham}, \citenamefont {Heisenberg},\ and\
  \citenamefont {Tolley}}]{Burrage:2011cr}%
  \BibitemOpen
  \bibfield  {author} {\bibinfo {author} {\bibfnamefont {C.}~\bibnamefont
  {Burrage}}, \bibinfo {author} {\bibfnamefont {C.}~\bibnamefont {de~Rham}},
  \bibinfo {author} {\bibfnamefont {L.}~\bibnamefont {Heisenberg}}, \ and\
  \bibinfo {author} {\bibfnamefont {A.~J.}\ \bibnamefont {Tolley}},\ }\href
  {\doibase 10.1088/1475-7516/2012/07/004} {\bibfield  {journal} {\bibinfo
  {journal} {JCAP}\ }\textbf {\bibinfo {volume} {1207}},\ \bibinfo {pages}
  {004} (\bibinfo {year} {2012})},\ \Eprint {http://arxiv.org/abs/1111.5549}
  {arXiv:1111.5549 [hep-th]} \BibitemShut {NoStop}%
\bibitem [{\citenamefont {Deser}\ and\ \citenamefont
  {Waldron}(2013)}]{Deser:2012qx}%
  \BibitemOpen
  \bibfield  {author} {\bibinfo {author} {\bibfnamefont {S.}~\bibnamefont
  {Deser}}\ and\ \bibinfo {author} {\bibfnamefont {A.}~\bibnamefont
  {Waldron}},\ }\href@noop {} {\bibfield  {journal} {\bibinfo  {journal} {Phys.
  Rev. Lett. 110,}\ }\textbf {\bibinfo {volume} {111101}} (\bibinfo {year}
  {2013})},\ \Eprint {http://arxiv.org/abs/1212.5835} {arXiv:1212.5835
  [hep-th]} \BibitemShut {NoStop}%
\bibitem [{\citenamefont {De~Felice}\ \emph {et~al.}(2013)\citenamefont
  {De~Felice}, \citenamefont {Gumrukcuoglu}, \citenamefont {Lin},\ and\
  \citenamefont {Mukohyama}}]{DeFelice:2013awa}%
  \BibitemOpen
  \bibfield  {author} {\bibinfo {author} {\bibfnamefont {A.}~\bibnamefont
  {De~Felice}}, \bibinfo {author} {\bibfnamefont {A.~E.}\ \bibnamefont
  {Gumrukcuoglu}}, \bibinfo {author} {\bibfnamefont {C.}~\bibnamefont {Lin}}, \
  and\ \bibinfo {author} {\bibfnamefont {S.}~\bibnamefont {Mukohyama}},\
  }\href@noop {} {\  (\bibinfo {year} {2013})},\ \Eprint
  {http://arxiv.org/abs/1303.4154} {arXiv:1303.4154 [hep-th]} \BibitemShut
  {NoStop}%
\bibitem [{\citenamefont {Izumi}\ and\ \citenamefont
  {Ong}(2013)}]{Izumi:2013poa}%
  \BibitemOpen
  \bibfield  {author} {\bibinfo {author} {\bibfnamefont {K.}~\bibnamefont
  {Izumi}}\ and\ \bibinfo {author} {\bibfnamefont {Y.~C.}\ \bibnamefont
  {Ong}},\ }\href@noop {} {\  (\bibinfo {year} {2013})},\ \Eprint
  {http://arxiv.org/abs/1304.0211} {arXiv:1304.0211 [hep-th]} \BibitemShut
  {NoStop}%
\end{thebibliography}%

\end{document}